\documentclass[preprint,proceedings]{rmaa}
\suppressfulladdresses
\usepackage{paralist}

\SetYear{2005}
\SetConfTitle{Latin American Regional IAU Meeting, Pucon}

\title{Understanding the Stellar Initial Mass Function}

\author{Richard B. Larson\altaffilmark{1}}

\altaffiltext{1}{Department of Astronomy, Yale University, New Haven, Connecticut, USA.}

\shortauthor{Richard B. Larson}
\shorttitle{Understanding the Stellar IMF}

\fulladdresses{\item Department of Astronomy, Yale University, P.O. Box 208101, New Haven, CT 06520-8101, USA (\email{richard.larson@yale.edu}).}

\listofauthors{R. B. Larson}
\indexauthor{Larson, R. B.}

\abstract{The essential features of the stellar Initial Mass Function are, rather generally, (1) a peak at a mass of a few tenths of a solar mass, and (2) a power-law tail toward higher masses that is similar to the original Salpeter function.  Recent work suggests that the IMF peak reflects a preferred scale of fragmentation associated with the transition from a cooling phase of collapse at low densities to a nearly isothermal phase at higher densities, where the gas becomes thermally coupled to the dust.  The Salpeter power law is plausibly produced, at least in part, by scale-free accretion processes that build up massive stars in dense environments.  The young stars at the Galactic Center appear to have unusually high masses, possibly because of a high minimum mass resulting from the high opacity of the dense star-forming gas.}

\addkeyword{Stars: Formation}
\addkeyword{Stars: Initial Mass Function}
\begin{document}
\maketitle

\section{Introduction}

   The year 2005 brought the 50th anniversary of studies of the stellar Initial Mass Function (IMF), which began with the publication of the first and still most famous paper on the subject by Salpeter in 1955.  Recent work on the subject has been summarized in the proceedings of a conference held to mark this occasion at Spineto, Italy (Corbelli, Palla, \& Zinnecker 2005), and useful reviews have been given also by Scalo (1986, 1998), Kroupa (2002), and Chabrier (2003).  Much work has shown that, for masses above 1\,M$_\odot$, the IMF can generally be approximated by a declining power law with a slope similar to that first found by Salpeter (1955).  However, it is now clear that this power law does not extend to masses much below 1\,M$_\odot$; instead, the number of stars per unit logarithmic mass interval flattens below 0.5\,M$_\odot$ and peaks at a few tenths of a solar mass, showing a significant decline in the brown-dwarf regime below 0.1\,M$_\odot$.  Thus, stars form with a preferred mass that is of the order of a few tenths of a solar mass.  If a characteristic mass is defined such that half of the initial mass goes into stars below this mass and half into stars above it, this characteristic stellar mass is about 1\,M$_\odot$.  This characteristic mass also seems to have a similar magnitude in many different well-studied systems.

   The fact that stars form with a preferred mass is perhaps the most fundamental fact about star formation, so I shall concentrate here on recent efforts to understand the origin of this basic mass scale.  Evidence for variations in this mass scale and their possible origin will be discussed briefly in Section~5.

\section{Origin of the Stellar Mass Scale}

   Studies of the structure of star-forming molecular clouds have suggested that low-mass stars may acquire their masses, at least in part, from those of the apparently prestellar dense `cores' observed in these clouds.  The mass spectrum of these cores resembles the stellar IMF for masses below a few M$_\odot$ (Motte, Andr\'e, \& Neri 1998; Testi \& Sargent 1998; Motte \& Andr\'e 2001; Johnstone et al.\ 2000, 2001), suggesting that low-mass stars may form directly from such cores and acquire their masses from them.  Although it is not yet clear to what extent there may be a one-to-one correspondence between cores and stars, the fact that the observed cores have masses similar to those of low-mass stars suggests that cloud fragmentation processes may play an important role in the origin of the IMF, especially in the origin of the characteristic stellar mass.

   Although turbulence and magnetic fields may dominate the dynamics of molecular clouds on the largest scales, thermal pressure is expected to be the most important force resisting gravity on the small scales relevant to the formation of individual stars.  Cloud fragmentation on these scales is therefore expected to be limited by the classical Jeans length and mass obtained by balancing gravity against thermal pressure (Jeans 1902, 1929).  Although the original Jeans analysis was not self-consistent, the Jeans mass or equivalent mass scales have been found to have wide applicability to many kinds of pressure-supported configurations (Larson 1985, 2003).  Many numerical simulations of cloud collapse and fragmentation have shown, furthermore, that the number of bound prestellar fragments that form is always similar to the number of Jeans masses in the initial configuration (Larson 1978; Monaghan \& Lattanzio 1991; Klessen, Burkert, \& Bate 1998; Klessen 2001; Bate, Bonnell, \& Bromm 2003; Bate \& Bonnell 2005).  The more recent of these simulations have mostly included a spectrum of turbulent motions, and they have found that the number of bound fragments formed is always similar to the number of Jeans masses present initially, regardless of whether turbulence is included or how it is treated.  Simulations including magnetic fields are still at an early stage, but the results of Li et al.\ (2004) suggest that magnetic fields, like turbulence, may not greatly alter the scale of fragmentation.

   If the characteristic stellar mass is therefore determined essentially by the initial Jeans mass in a collapsing cloud, we need to understand what determines this initial Jeans mass, or the Jeans mass at the stage when most of the fragmentation takes place.  The Jeans mass depends on the gas temperature and density (or pressure), but it is particularly sensitive to the temperature, varying as temperature to the power 3/2 or 2 depending on whether the density or the pressure is fixed.  This suggests that the thermal properties of star-forming clouds are important for the scale of fragmentation, and that any variation in temperature during the collapse may also have important effects.  The simulations mentioned above have nearly all assumed isothermal collapse, but isothermality is only a somewhat crude approximation and the cloud temperature is actually predicted to drop at first with increasing density because of efficient cooling at low densities, reaching a minimum and then beginning to rise slowly with density when the gas becomes thermally coupled to the dust (Larson 1985).  Fragmentation might then be expected to proceed very efficiently at the low densities where cooling is efficient and the Jeans mass decreases rapidly with increasing density, while it might be much less efficient at the higher densities where the Jeans mass decreases only slowly with increasing density.  Larson (1985) suggested on this basis that the Jeans mass at the point of minimum temperature where the gas becomes thermally coupled to the dust might be a preferred scale of fragmentation, possibly leading to a peak in the stellar IMF at this mass, which is predicted to be about 0.3\,M$_\odot$, in agreement with the observed peak mass.  A similar suggestion that the peak of the IMF is determined by the onset of thermal coupling between gas and dust was made by Whitworth, Boffin, \& Francis (1998).

\section{Importance of the Thermal Physics of Star-Forming Clouds}

   Recent numerical experiments have shown that the dependence of fragmentation efficiency on thermal behavior suggested above is in fact a very strong effect.  Li, Mac Low, \& Klessen (2003) have simulated the collapse and fragmentation of a turbulent cloud with a polytropic equation of state of the form $P \propto \rho^{\gamma}$ and have considered a range of values of $\gamma$.  They found that the number of bound clumps formed decreases very strongly with increasing $\gamma$; for example, it decreases from 380 for $\gamma = 0.7$ to 18 for $\gamma = 1.1$, a decrease by a factor of 20 in the number of bound clumps formed for only a modest decrease in $\gamma$.  This range of values of $\gamma$ is approximately the range expected to be relevant in star-forming clouds, as was noted by Larson (1985).  Studies of the thermal behavior of star-forming clouds have been reviewed by Larson (2005), including the work of Larson (1973), Low \& Lynden-Bell (1976), Koyama \& Inutsuka (2000), Masunaga \& Inutsuka (2000), and Omukai (2000).  At densities below about $10^{-18}$\,g\,cm$^{-3}$, the balance between radiative heating and cooling mainly by C$^+$ ions and O atoms results in a temperature that decreases with increasing density roughly as $T=4.4\,\rho_{18}^{-0.27}\,{\rm K}$, where $\rho_{18}$ is the density in units of $10^{-18}$\,g\,cm$^{-3}$.  At higher densities, the gas becomes thermally coupled to the dust and the balance between compressional heating and dust cooling yields a slowly rising temperature that can be approximated by $T=4.4\,\rho_{18}^{0.07}\,{\rm K}$.  Thus the effective value of $\gamma$ increases from about 0.73 to 1.07 at the density of about $10^{-18}$\,g\,cm$^{-3}$ where the gas becomes thermally coupled to the dust.  From the results of Li et al.\ (2003) we would then expect that efficient fragmentation would occur at the lower densities where $\gamma \simeq 0.73$, while relatively little further fragmentation would occur at the higher densities where $\gamma \simeq 1.07$, making the Jeans mass at the transition density a preferred scale for fragmentation (Larson 2005).

   The hypothesis that the Jeans mass at the density where the gas becomes thermally coupled to the dust is what determines the peak mass of the IMF has been tested in a series of calculations by Jappsen et al. (2005), who simulated the fragmentation of a cloud with a two-part polytropic equation of state in which $\gamma$ increases from 0.7 to 1.1 at a critical density which is varied to study its effect on the mass spectrum of the objects formed.  These simulations produced a few hundred bound objects and thus yielded statistically meaningful mass functions.  The results showed, at least qualitatively, the expected strong dependence of the peak mass of the IMF on the critical density, confirming the importance of the thermal physics of star-forming clouds in determining the mass scale of the IMF.  The simulation with the most realistic choice of parameters yielded an IMF with a peak mass of about 0.3\,M$_\odot$, very similar to the observed value.  Further simulations testing the importance of the thermal physics have been made by Bonnell, Clarke, \& Bate (2006a), who found that while a calculation with an isothermal equation of state yielded an acceptable IMF only if the right initial conditions were chosen, one with a more realistic equation of state like that discussed above yielded an IMF similar to the observed one with a flattening below 1\,M$_\odot$ and a peak at a few tenths of a solar mass, even when the initial Jeans mass was much larger than 1\,M$_\odot$.  Again, this result supports the hypothesis that the change in the equation of state that occurs when the gas becomes thermally coupled to the dust is what is responsible for determining the characteristic stellar mass.

\section{Massive Stars and the Upper IMF}

   Fragmentation processes may account for the lower IMF, i.e.\ for stars with masses below a few M$_\odot$, but they are not likely to produce the more massive stars or account for the upper IMF because the most massive stars form in dense environments where the Jeans mass is small, typically in the central parts of large dense clusters like the Trapezium cluster (Zinnecker, McCaughrean, \& Wilking 1993).  The tendency of massive stars to form in dense regions where accumulation processes are likely to be significant suggests instead that the upper IMF may be produced, at least in part, by the continuing accretional growth of massive stars in dense environments (Larson 1982; Zinnecker 1982; Bonnell, Bate, \& Vine 2003; Bonnell, Larson, \& Zinnecker 2006b).  Because the stars in a forming cluster can all accrete from a common intracluster medium, the term `competitive accretion' has sometimes been used to describe this process (Bonnell et al.\ 2001).  The simplest approximation, assuming Bondi-Hoyle accretion at a rate proportional to the square of the mass of the accreting object, predicts the growth of a power-law tail on the IMF with a logarithmic slope of $-1$ which is not far from the Salpeter slope of $-1.35$ (Zinnecker 1982).  The tendency of the more massive stars to be located in denser regions creates a stronger dependence of the accretion rate on stellar mass, and this increases the slope of the IMF to a value closer to the Salpeter slope.  Simulations of the formation of clusters have produced not only realistic-looking clusters but also a realistic upper stellar IMF that has a slope similar to the Salpeter slope (Bonnell et al.\ 2003, 2006b).  These results support the possibility that the upper IMF results largely from accretion processes, although the details of the accretion processes that occur in the simulations are much more complex than classical Bondi-Hoyle accretion because the accreted material is highly non-uniform and because it has very complex dynamics.

   A possible problem for models in which accretion plays an important role is posed by the effect of radiation pressure on the gas falling onto an accreting star; if the star becomes too luminous, its radiation pressure may shut off further growth in mass at a stellar mass of a few tens of solar masses (Larson \& Starrfield 1971; Wolfire \& Cassinelli 1987; Yorke \& Sonnhalter 2002).  Recent detailed three-dimensional calculations of the effects of radiation on an infalling envelope have shown, however, that radiation pressure is not a serious obstacle to continuing accretion because Rayleigh-Taylor instabilities or the formation of bipolar outflow cavities can allow radiation to escape and infall to continue and build up stars much more massive than would otherwise be predicted (Krumholz, McKee, \& Klein 2005a, 2006).  Although Krumholz et al.\ (2005b) have argued against the importance of competitive accretion for the formation of massive stars, their work on radiation pressure actually supports the possibility of building up massive stars by accretion processes of some kind.  Simple models cannot be used to argue for or against competitive accretion because of the extreme density contrasts and complex dynamics involved, as is found in the simulations.  In any case, much clearly remains to be learned about the formation of massive stars.

\section{Star Formation in Extreme Environments}

   Most well-studied systems appear to have a similar IMF, but evidence has recently begun to emerge that the Galactic Center region may have an anomalous IMF characterized by an excess of massive stars.  One of the three massive young clusters near the Galactic Center, the Arches Cluster, has a relatively flat upper IMF with an apparent turnover at a few solar masses, suggesting a characteristic mass considerably higher than in the solar neighborhood (Stolte et al.\ 2005).  A second young cluster, the Galactic Center cluster surrounding the central black hole, appears to be deficient in low-mass stars by more than  an order of magnitude, judging by the absence of the predicted X-ray flux from the low-mass pre-main-sequence stars that should be present if the cluster has a normal IMF (Nayakshin \& Sunyaev 2005).  The K-band luminosity function of this cluster also indicates a top-heavy IMF, and from all of the evidence, it seems very likely that these stars were formed essentially where they are, within a few tenths of a parsec of the central black hole (Paumard et al.\ 2006; Nayakshin \& Sunyaev 2005).  Thus, there is now good evidence that stars have recently formed very close to the central black hole in our Galaxy with a very top-heavy IMF (Genzel 2006).

   In order for stars to form so close to the central black hole, the star-forming gas must have a density many orders of magnitude higher than that of typical nearby molecular clouds for it not to be dispersed by tidal forces.  Most of the stars in the Galactic Center cluster are concentrated in two nearly orthogonal disks or rings (Genzel et al.\ 2003; Paumard et al.\ 2006), and these stars must presumably have formed in two dense disks of gas orbiting the central black hole (Nayakshin \& Cuadra 2005; Nayakshin 2006).  The gas in these star-forming disks must have had a density of at least $10^{-13}$\,g\,cm$^{-3}$, which is high enough for the gas to be closely thermally coupled to the dust.  Moreover, at this high density any region large enough to be self-gravitating is optically thick to the thermal cooling radiation of the dust, and it may therefore tend to evolve adiabatically.  Because the dust opacity increases with temperature and because the dust in the Galactic Center region is much warmer than the dust in nearby molecular clouds (Morris \& Serabyn 1996), the dust opacity will in fact be so high that any self-gravitating region must already have become opaque at a much lower density.  This optically thick regime is one that has not previously received much attention in star formation theory, except in the context of the `opacity limit' on stellar masses first studied by Low \& Lynden-Bell (1976).

   If the dust is externally heated, as appears to be the case near the Galactic Center, the minimum fragment mass predicted by Low \& Lynden-Bell (1976) increases strongly with the dust temperature and may reach values much higher than the normal value.  If the dust opacity varies as $T_{\rm dust}^2$, as expected, and if the gas behaves adiabatically after it becomes opaque, as assumed by Low \& Lynden-Bell, then the minimum fragment mass is predicted to vary as $T_{\rm dust}^4$.  Dust temperatures exceeding 50\,K are observed near the Galactic Center, and for temperatures in this range the predicted minimum fragment mass becomes larger than 1\,M$_\odot$.  Since the dust temperature may have been even higher than this during the period of active star formation that created the observed massive stars, the minimum mass could have been even higher, implying a top-heavy IMF, as observed.  Fragmentation to smaller masses could have occurred only if the gas in the star-forming disks was able to cool to lower temperatures before forming stars.  For example, if the gas was able to cool to a constant temperature of 40\,K, as suggested by Nayakshin (2006), then the formation of stars with masses below 1\,M$_\odot$ might have been possible, but even in this case, Nayakshin (2006) argues that these stars would have grown rapidly by accretion to much larger masses, so that the end result would still have been massive stars. 

   Clearly the above discussion is very preliminary, and new frontiers remain to be explored in the study of star formation in extreme environments like that near the Galactic center.  Another type of extreme environment that has been studied is that relevant to the formation of the first stars in the universe; these are also predicted to be relatively massive because of the expected high temperature and high mass scale for fragmentation (Bromm \& Larson 2004).  A subject not yet well studied is the formation of the first very metal-poor stars after the first heavy elements had been produced in the universe.  With a low dust abundance, the resulting low opacity may allow the formation of low-mass stars even in very metal-poor systems, as long as even a small amount of dust is present (Low \& Lynden-Bell 1976; Omukai et al.\ 2005); this might help to explain why the IMF of very metal-poor systems does not appear to be markedly anomalous.  Detailed calculations will be needed to understand what kind of IMF might result from such conditions.  Perhaps, with enough work and with the efforts of enterprising students like those attending this meeting, an understanding of the origin of the IMF in the various different astrophysical environments relevant to the formation and evolution of galaxies will eventually become possible.


\begin{thebibliography}

\bibitem{}Bate, M. R., Bonnell, I. A., \& Bromm, V. 2003, MNRAS, 339, 537

\bibitem{}Bate, M. R., \& Bonnell, I. A. 2005, MNRAS, 356, 1201

\bibitem{}Bonnell, I. A., \& Bate, M. R., 2002, MNRAS, 336, 659

\bibitem{}Bonnell, I. A., Bate, M. R., Clarke, C. J., \& Pringle, J. E. 2001,
  MNRAS, 323, 785

\bibitem{}Bonnell, I. A., Bate, M. R., \& Vine, S. G. 2003, MNRAS, 343, 413

\bibitem{}Bonnell, I. A., Clarke, C. J., \& Bate, M. R. 2006a, MNRAS, in press
  (astro-ph/0603444)
  
\bibitem{}Bonnell, I. A., Larson, R. B., \& Zinnecker, H. 2006b, in Protostars
  and Planets V, eds. B. Reipurth, D. C. Jewitt, \& K. Keil (Univ.\ of
  Arizona Press, Tucson), in press (astro-ph/0603447)

\bibitem{}Chabrier, G. 2003, PASP, 115, 763

\bibitem{}Corbelli, E., Palla, F., \& Zinnecker, H. 2005, The Initial Mass
  Function 50 Years Later (Springer, Dordrecht)

\bibitem{}Genzel, R., 2006, in 11th Latin American Regional IAU Meeting, Pucon,
  Chile, Rev.\ Mex.\ Astr.\ Ap., in press (this conference)

\bibitem{}Genzel, R., Sch\"odel, R., Ott, T., Eisenhauer, F., Hoffman,
  R., Lehnert, M., Eckart, A., Alexander, T., Sternberg, A., Lenzen, R.,
  Cl\'enet, Y., Lacombe, F., Rouan, D., Renzini, A., \& Tacconi-Garman,
  L. E. 2003, ApJ, 594, 812

\bibitem{}Jappsen, A-K., Klessen, R. S., Larson, R. B., Li, Y., \& Mac Low,
  M.-M., 2005, A\&A, 435, 611

\bibitem{}Jeans, J. H. 1902, Phil.\ Trans.\ Roy.\ Soc.\ A, 199, 49

\bibitem{}Jeans, J. H. 1929, Astronomy and Cosmogony (Cambridge Univ.\
  Press, Cambridge)

\bibitem{}Johnstone, D., Wilson, C. D., Moriarty-Schieven, G., Joncas, G.,
  Smith, G., Gregersen, E., \& Fich, M., 2000, ApJ, 545, 327

\bibitem{}Johnstone, D., Fich, M., Mitchell, G. F., \& Moriarty-Schieven,
  G., 2001, ApJ, 559, 307

\bibitem{}Klessen, R. S. 2001, ApJ, 556, 837

\bibitem{}Klessen, R. S., Burkert, A., \& Bate, M. R. 1998, ApJ, 501, L205

\bibitem{}Koyama, H., \& Inutsuka, S.-I., 2000, ApJ, 532, 980

\bibitem{}Kroupa, P. 2002, Science, 295, 82

\bibitem{}Krumholz, M. R., McKee, C. F., \& Klein, R. I. 2005a, ApJ, 618, L33

\bibitem{}Krumholz, M. R., McKee, C. F., \& Klein, R. I. 2005b, Nature, 438, 332

\bibitem{}Krumholz, M. R., McKee, C. F., \& Klein, R. I. 2006, in IAU Symp.\
  227, Massive Star Birth: A Crossroads of Astrophysics, eds.\ R. Cesaroni,
  E. Churchwell, M. Felli, \& C. M. Walmsley (Cambridge Univ.\ Press,
  Cambridge), in press (astro-ph/0510432)

\bibitem{}Larson, R. B. 1973, Fundam.\ Cosmic Phys., 1, 1

\bibitem{}Larson, R. B. 1978, MNRAS, 184, 69

\bibitem{}Larson, R. B. 1982, MNRAS, 200, 159

\bibitem{}Larson, R. B. 1985, MNRAS, 214, 379

\bibitem{}Larson, R. B. 2003, Rep.\ Prog.\ Phys., 66, 1651

\bibitem{}Larson, R. B. 2005, MNRAS, 359, 211

\bibitem{}Larson, R. B., \& Starrfield, S. 1971, A\&A, 13, 190

\bibitem{}Li, P. S., Norman, M. L., Mac Low, \& Heitsch, 2004, ApJ, 605, 800

\bibitem{}Li, Y., Klessen, R. S., \& Mac Low, M.-M., 2003, ApJ, 592, 975

\bibitem{}Low, C., \& Lynden-Bell, D. 1976, MNRAS, 176, 367

\bibitem{}Mac Low, M.-M., \& Klessen R. F. 2004, Rev.\ Mod.\ Phys., 76, 125

\bibitem{}Masunaga, H., \& Inutsuka, S.-I., 2000, ApJ, 531, 350

\bibitem{}Monaghan, J. J., \& Lattanzio, J. C. 1991, ApJ, 375, 177

\bibitem{}Morris, M., \& Serabyn, E. 1996, ARA\&A, 34, 645

\bibitem{}Motte, F., \& Andr\'e, P. 2001, in ASP Conf.\ Ser.\ 243, From
  Darkness to Light: Origin and Evolution of Young Stellar Clusters, eds.\
  T. Montmerle \& P. Andr\'e (Astr.\ Soc.\ Pacific, San Francisco), p.~301 

\bibitem{}Motte, F., Andr\'e, P., \& Neri, R. 1998, A\&A, 336, 150

\bibitem{}Nayakshin, S. 2006, MNRAS, in press (astro-ph/0512255)

\bibitem{}Nayakshin, S., \& Cuadra, J. 2005, A\&A, 437, 437

\bibitem{}Nayakshin, S., \& Sunyaev, R. 2005, MNRAS, 364, L23

\bibitem{}Omukai, K., 2000, ApJ, 534, 809

\bibitem{}Omukai, K., Tsuribe, T., Schneider, R., \& Ferrara, A. 2005, ApJ,
  626, 627 

\bibitem{}Paumard, T., Genzel, R., Martins, F., Nayakshin, S., Beloborodov,
  A. M., Levin, Y., Trippe, S., Eisenhauer, F., Ott, T., Gillessen, S.,
  Abuter, R., Cuadra, J., \& Alexander, T. 2006, ApJ, in press
  (astro-ph/0601268)

\bibitem{}Salpeter, E. E. 1955, ApJ, 121, 161

\bibitem{}Scalo, J. M. 1986, Fundam.\ Cosmic Phys., 11, 1

\bibitem{}Scalo, J. M. 1998, in ASP Conf.\ Ser.\ 142, The Stellar Initial
  Mass Function, eds.\ G. Gilmore \& D. Howell (Astr.\ Soc.\ Pacific, San
  Francisco), p.~201  

\bibitem{}Stolte, A., Brandner, W., Grebel, E. K., Lenzen, R., \& Lagrange,
  A.-M., 2005, ApJ, 628, L113

\bibitem{}Testi, L., \& Sargent, A. I. 1998, ApJ, 508, L91

\bibitem{}Whitworth, A. P., Boffin, H. M. J., \& Francis, N. 1998, MNRAS,
  299, 554

\bibitem{}Wolfire, M. G., \& Cassinelli, J. P. 1987, ApJ, 319, 850 

\bibitem{}Zinnecker, H. 1982, in Symposium on the Orion Nebula to Honor
  Henry Draper, eds.\ A. E. Glassgold, P. J. Huggins, \& E. L. Schucking
  (Ann.\ New York Acad.\ Sci., 395, New York), p.~226

\bibitem{}Zinnecker, H., McCaughrean, M. J., \& Wilking, B. A. 1993, in
  Protostars and Planets III, eds.\ E. H. Levy \& J. I. Lunine (Univ.\
  of Arizona Press, Tucson), p.~429

 
\end{thebibliography}
\end{document}